\pdfoutput=1
\documentclass[aps,prb,twocolumn,superscriptaddress]{revtex4-1}
\usepackage{graphicx}
\newcommand{\be}{\begin{equation}}
\newcommand{\ee}{\end{equation}}
\newcommand{\bea}{\begin{eqnarray}}
\newcommand{\eea}{\end{eqnarray}}
\newcommand{\gc}{{\gamma_{\rm C}}}

\newcommand{\gxone}{{\gamma_{\rm X1}}}
\newcommand{\gxtwo}{{\gamma_{\rm X2}}}
\newcommand{\gxthree}{{\gamma_{\rm X3}}}

\newcommand{\wc}{{\omega_{\rm C}}}
\newcommand{\wxn}{{\omega_{{\rm X}\!n}}}
\newcommand{\wxone}{{\omega_{\rm X1}}}
\newcommand{\wxtwo}{{\omega_{\rm X2}}}
\newcommand{\wxthree}{{\omega_{\rm X3}}}

\usepackage{ifpdf}

\newif\iffigs

\figstrue

\usepackage{todonotes}

\begin{document}

%\preprint{}

\title{Microcavity controlled coupling of excitonic qubits}



\author{F. Albert}
\affiliation{Technische Physik, Physikalisches Institut, and Wilhelm Conrad R\"ontgen Research
Center for Complex Material Systems, Universit\"at W\"urzburg, Am Hubland, D-97074 W\"urzburg,
Germany}
\author{K. Sivalertporn}
\affiliation{Cardiff University School of Physics and Astronomy, The Parade, Cardiff CF24 3AA,
United Kingdom}
\author{J. Kasprzak}
\affiliation{Institut N\'eel, CNRS et Universit\'e Joseph Fourier, BP 166, F-38042 Grenoble Cedex
9, France}
\author{M Strau\ss}
\author{C. Schneider}
\author{S. H\"ofling}
\author{M. Kamp}
\author{A. Forchel}
\affiliation{Technische Physik, Physikalisches Institut, and Wilhelm Conrad R\"ontgen Research
Center for Complex Material Systems, Universit\"at W\"urzburg, Am Hubland, D-97074 W\"urzburg,
Germany}
\author{S. Reitzenstein}
\altaffiliation[Present Address: ] {Institut f\"ur Festk\"orperphysik, Technische Universit\"at
Berlin, Hardenbergstrasse 36, 10623 Berlin, Germany}\affiliation{Technische Physik, Physikalisches
Institut, and Wilhelm Conrad R\"ontgen Research Center for Complex Material Systems, Universit\"at
W\"urzburg, Am Hubland, D-97074 W\"urzburg, Germany}
\author{E. A. Muljarov}
\altaffiliation{On leave from General Physics Institute RAS, Moscow, Russia}
\author{W. Langbein}
\affiliation{Cardiff University School of Physics and Astronomy, The Parade, Cardiff CF24 3AA,
United Kingdom}
\email{langbeinww@cardiff.ac.uk}

\begin{abstract} Controlled non-local energy and coherence transfer enables light harvesting in
photosynthesis and non-local logical operations in quantum computing. The most relevant mechanism
of coherent coupling of distant qubits is coupling via the electromagnetic field. Here, we
demonstrate the controlled coherent coupling of spatially separated excitonic qubits via the photon
mode of a solid state microresonator. This is revealed by two-dimensional spectroscopy of the
sample's coherent response, a sensitive and selective probe of the coherent coupling. The
experimental results are quantitatively described by a rigorous theory of the cavity mediated
coupling within a cluster of quantum dots excitons. Having demonstrated this mechanism, it can be
used in extended coupling channels - sculptured, for instance, in photonic crystal cavities - to
enable a long-range, non-local wiring up of individual emitters in solids.
\end{abstract}


\date{\today}

%\pacs{}

\maketitle

Sunlight absorption via antenna proteins and the subsequent resonant energy transfer over a few
nanometers towards the reaction centre is at the heart of the photosynthesis process
\cite{ColliniNature10,ScholesNatureChem11}. In this case the electronic or F\"orster coupling
strength and the local vibrational dissipation is tuned to achieve a fast directional energy
transfer. Instead, in quantum information science and cavity quantum electrodynamics (cQED), the
quantum bus technology aims to provide dissipation-less coupling between distant quantum systems.
Within this field, long-range coherent coupling between individual, distant superconducting qubits
has recently been demonstrated \cite{MajerNature07}, so that the construction of quantum logic
gates and networks is within the reach of present technology \cite{SchoelkopfNature08, LeekPRB09,
ChowPRL11}. In this context, qubits embedded in a solid state matrix are attractive, as they can
fully benefit from the lithographic and materials processing techniques developed in the
semiconductor industry to fabricate on-demand photonic layouts, so as to enable long-range coherent
coupling within a discrete set of qubits via a photonic intra-cavity bus. Moreover, their ``niche"
with respect to superconducting qubits is marked out by a simultaneous functionality at relatively
high temperatures and at optical frequencies, and they provide an interface to flying qubits such
as photons in optical fibres.

\begin{figure*}
\iffigs\ifpdf\includegraphics*[width=0.9\textwidth]{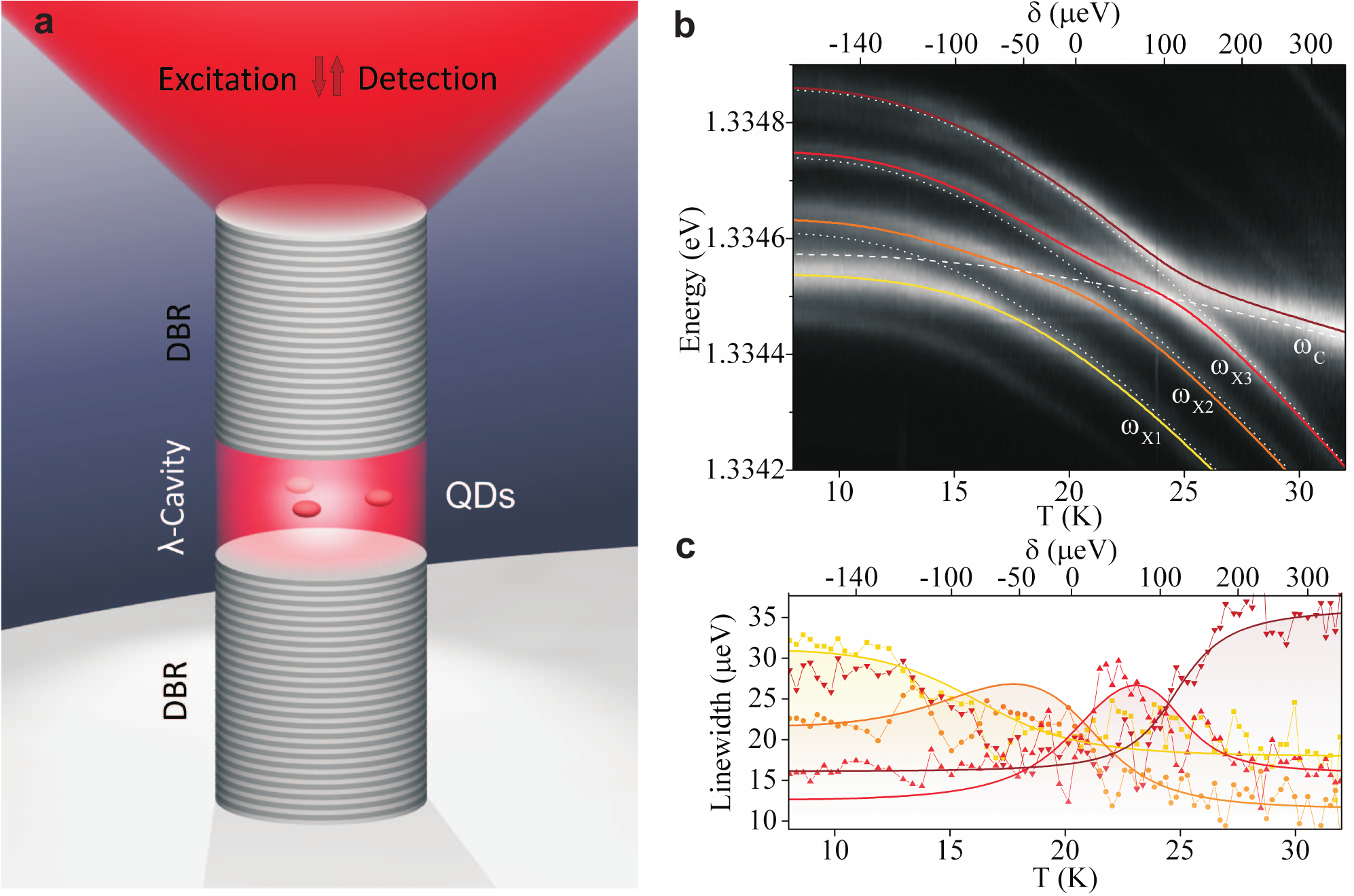}\fi\fi \caption{ Characterization of
the investigated quantum dot - micropillar system. a) Sketch of the micropillar structure including
the light coupling from the top facet. b) Temperature dependent photoluminescence spectral
intensity under non-resonant excitation on a linear grey scale black (0) to white. The bare
resonance energies of excitons and the cavity mode (white dotted and dashed lines, respectively),
and the coupled polariton energies (solid lines) obtained from a Lorentzian lineshape fit and
modelling (see supplementary material) are overlayed to the data. The corresponding average
detuning $\delta$ (see Eqn.\,\ref{eqn:detuning}) is shown on the upper axis scale. c) Coupled
resonance linewidths (measured: symbols, modeling: lines). Colours and linestyles as in b).
\label{fig:microPL}}
\end{figure*}

In recent years, significant progress has been made in the realization of high quality optical
microresonators which enabled pioneering demonstrations of the strong \cite{ReithmaierNature04,
YoshieNature04,PeterPRL05,OtaAPL11} and quantum strong coupling regime \cite{FaraonNaturePhys08,
KasprzakNatureMat10} in solid state. Micropillar cavities (cf. Fig.\,\ref{fig:microPL}a) are a
model system for the study of strong coupling in this field.  They consist of self-assembled InGaAs
quantum dots (QD) - providing individual exciton states (X) of high oscillator strength, located in
the anti-node of the fundamental cavity mode (C). In these structures a quantum of optical
excitation coherently oscillates between the fermionic exciton and bosonic cavity photon state. The
resulting eigenstates of mixed exciton and photon character form a Jaynes-Cummings (JC)
ladder\cite{JaynesProcIEEE63} with increasing number of photons in the cavity mode, showing a Rabi
splitting of the rungs proportional to the root of the photon number.

 In this Letter we report on coherent measurements and modeling of cavity-mediated
coherent coupling between three quantum dot excitons, moving from the JC ladder to the
Tavis-Cummings (TC) ladder \cite{TavisPR68}. This constitutes a crucial step towards a quantum bus
based on semiconductor photonic structures.

Coherent photonic coupling of distant qubits is realized here by their dipole interaction with a
common optical mode of a high-quality microresonator.  For this purpose we have chosen a
micropillar similar to that employed in Ref.\,\onlinecite{KasprzakNatureMat10} as described in the
Methods section. By micro-photoluminescence ($\mu$PL) measurements as shown in
Fig.\,\ref{fig:microPL}b, we have identified a triplet of Xs - labeled X1, X2 and X3, which at a
temperature of 8\,K are slightly blue-shifted from the cavity. Increasing the sample temperature,
the Xs (short-dashed lines) are tuned through C (long-dashed line) due to a reduction of the
semiconductor band-gap. The data reveal three X-C avoided crossings at around 13\,K, 21\,K and
25\,K, showing that each of them is in the strong coupling regime.

\begin{figure}
\iffigs\ifpdf\includegraphics*[width=\columnwidth]{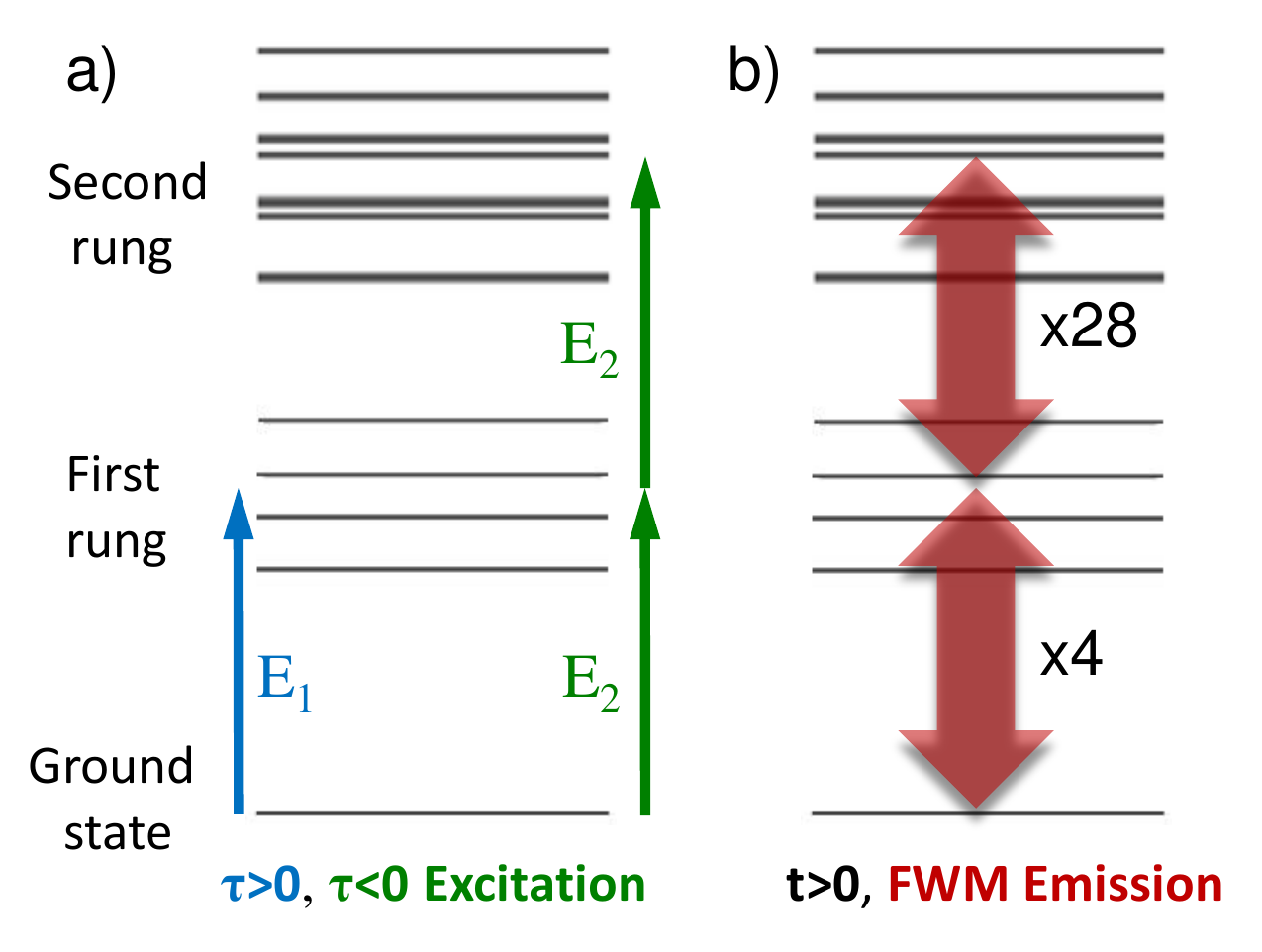}\fi\fi \caption{Level scheme of the
Tavis-Cummings ladder of the three exciton - one cavity system, and transitions relevant for the
coherent FWM response, for $\delta = -29\,\mu$eV. a) Coherence created by the pulse arriving first
($E_1$ for $\tau>0$, $E_2$ for $\tau<0$). b) Transitions emitting FWM after the arrival of the
second pulse.\label{fig:levels}}
\end{figure}

\begin{figure*} \iffigs\ifpdf\includegraphics*[width=\textwidth]{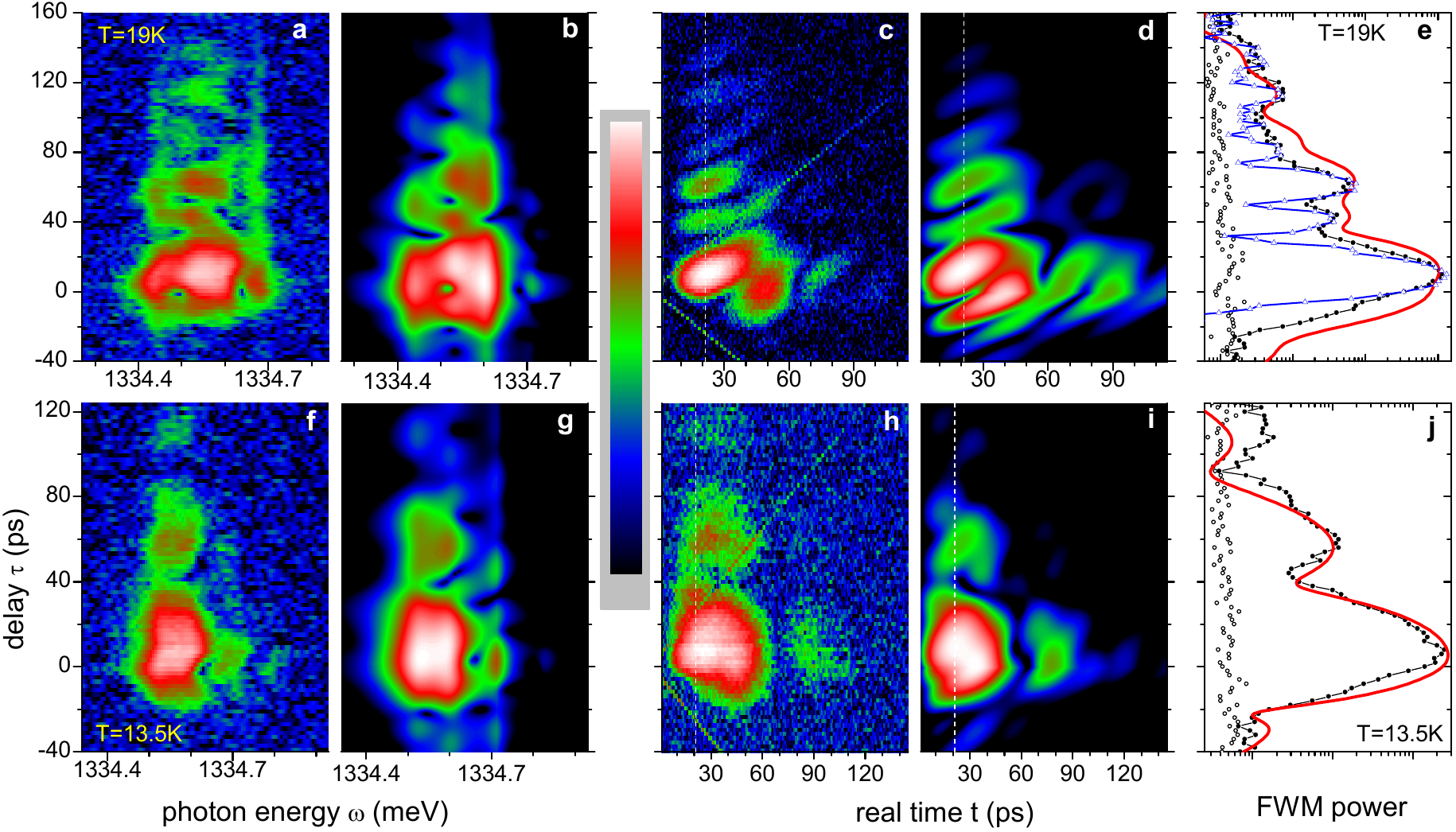}\fi\fi
\caption{\label{fig:FWMDynamics} Delay time dependence of coherent response for $T=19\,$K (top) and
$T=13.5\,$K (bottom). Spectrally resolved FWM power $|\tilde{P}(\omega,\tau)|^2$, measured (a,f)
and predicted (b,g), on a logarithmic colour scale over 4 orders of magnitude. Time-resolved FWM
power $|P(t,\tau)|^2$, measured (c,h) and predicted (d,i), over 3 orders. (e,j) Time-integrated FWM
power $|P|^2_{\rm int}(\tau)$, measured (black circles) and predicted (red line), and measured
$|P(21\,{\rm ps},\tau)|^2$ (blue triangles). The noise of $|P|^2_{\rm int}(\tau)$ is given as open
circles.} \end{figure*}

A triple exciton - cavity system has a level scheme as shown in Fig.\,\ref{fig:levels}, more
complex than the previously studied single exciton-cavity system \cite{KasprzakNatureMat10}. It
hosts four polaritonic transitions from the vacuum state. The polariton frequency tuning (solid
lines in Fig.\,\ref{fig:microPL}b) as well as the variation of polariton linewidths (solid lines in
Fig.\,\ref{fig:microPL}c) can be described by a coupled oscillator model with the X1-C, X2-C, X3-C
coupling parameters $(g_1, g_2, g_3)=(43, 40, 31.5) \,\mu$eV, homogeneous broadenings $(\gxone,
\gxtwo, \gxthree, \gc)=(18, 11.5, 16, 36.5)\,\mu$eV, and frequency distances
$\wxtwo-\wxone=131\,\mu$eV, $\wxthree-\wxone=248\,\mu$eV. The parameters were obtained from a
global fit of the coupled oscillator model to the detuning-dependent transition energies and
broadenings determined by Lorentzian lineshape fitting of the $\mu$PL-spectra in
Fig.\,\ref{fig:microPL}b as described in the supplementary material (SM). To describe the detuning
in this TC system with non-identical two-level systems, we introduce the average cavity detuning
\be\delta=\wc-\left.{\sum\limits_{n=1}^3  g_n \wxn } \right/ { \sum\limits_{n=1}^3
g_n}\label{eqn:detuning}\ee
having the values of $\delta = (-124,16,124)\,\mu$eV at the three anti-crossing points and shown in
the top axis of Fig.\,\ref{fig:microPL}b. Multi-polaritonic features as reported in
Fig.\,\ref{fig:microPL}b,c were previously observed in $\mu$PL experiments \cite{ReitzensteinOL06,
LauchtPRB10} and indicate via an anti-crossing behaviour between individual excitons and the cavity
mode that Xs could be coherently wired up by the cavity field. However, a direct measurement of
coherent coupling is not afforded by $\mu$PL measuring the incoherent emission, neither does it
provide means to perform coherent optical manipulation in prospective quantum bus nanophotonic
structures. An explicit demonstration and selective manipulation of the cavity-mediated coherent
coupling of excitons requires extracting and controlling the coherent response of the photonic
resonator.

To demonstrate and study the cavity mediated coherent coupling of Xs we employ here heterodyne
spectral interferometry \cite{LangbeinOL06} (HSI) to detect the cavity-mediated coherent coupling
between the excitons. This technique is used to measure the four-wave mixing (FWM) of the
strongly-coupled Xs-C system, which arises from the optical nonlinearity of the quantum dot
excitons. In short, two pulses $E_1$ and $E_2$ of 1\,ps duration and variable delay $\tau$ are
exciting the fundamental cavity mode, see Fig.\,S6 in the SM. The resulting FWM polarization
emitted from the micropillar $P(t,\tau) \propto E_1^*E_2^2$ and its Fourier transform (FT) versus
$t$, $\tilde{P}(\omega,\tau)$, are measured using spectral interferometry.

The coherent dynamics giving rise to FWM involves the four Xs-C mixed states of the first rung of
the TC ladder, which were identified in $\mu$PL, and additionally the 7 polariton states of the
second rung which are mixed states of the family of uncoupled exciton-photon states of the second
rung: one two-photon state without excitons, three one-photon states with one of the three excitons
filled, and three zero-photon states with two of the three exciton states filled. A sketch of the
resulting ladder of levels at T=19\,K ($\delta = -29\,\mu$eV) is shown in Fig.\,\ref{fig:levels},
together with the transitions relevant for the delay-time and real-time coherent dynamics probed in
FWM. For positive delay $\tau>0$, pulse $E_1$ arrives first and creates a one-photon coherence
given by a wavepacket of the 4 states of the first rung, which is coherently evolving until the
arrival of $E_2$. Conversely, for negative delay $\tau<0$, pulse $E_2$ arrives first and creates a
two-photon coherence given by a wavepacket in the second rung, which is coherently evolving until
the arrival of $E_1$. In both cases, at the arrival of the second pulse FWM is created as a
superposition of all optical transitions between the ground state, first and second rung,
consisting of 4 transitions between the ground state and the first rung and 28 transitions between
the first and second rung of the TC ladder. We calculate the FWM polarization analytically taking
into account the states up to the second rung, by solving the master equation for the density
matrix using a standard Xs-C coupling Hamiltonian and a Lindblad dissipation operator (see SM).
This approach is exact for the third-order FWM signal of an initially unexcited system.

The resulting measurements and corresponding predictions of FWM of the system are given in
Fig.\,\ref{fig:FWMDynamics} as function of time delay $\tau$ for two different detuning parameters.
For $\delta=-29\,\mu$eV ($T=19\,$K), the spectrally resolved $|\tilde{P}(\omega,\tau)|^2$ and
time-resolved $|P(t,\tau)|^2$ are shown in Figs.\,\ref{fig:FWMDynamics}a,c, respectively. A
dynamics significantly richer than in a single exciton case \cite{KasprzakNatureMat10} is observed,
as expected from the larger number of levels in the first and second rungs, providing 32 instead of
6 transitions contributing to the FWM (see Fig.\,\ref{fig:levels}). The time-integrated FWM power
$|P|^2_{\rm int}(\tau)=\int|P(t,\tau)|^2 dt$ and the power $|P(t_{\rm m},\tau)|^2$ at a given time
$t_{\rm m}=21$\,ps corresponding to the build-up lag of the FWM in such strongly coupled
exciton-cavity systems \cite{KasprzakNatureMat10} are presented in Fig.\,\ref{fig:FWMDynamics}e. On
a qualitative level, we notice the FWM beat as a function of $\tau$ with a period of about 17\,ps,
corresponding to a spectral splitting of $243\,\mu$eV. This is much larger than the Rabi splitting
of any individual X, and is close to the total splitting of $2(g_1+g_2+g_3)=229\,\mu$eV, indicating
that all four polaritons contribute towards the coherent dynamics. In
Figs.\,\ref{fig:FWMDynamics}b,d we present the predicted FWM corresponding to
\ref{fig:FWMDynamics}a,c, using the exciton and cavity parameters retrieved from the $\mu$PL data
(see Fig.\,\ref{fig:microPL} and the SM). The prediction, which takes into account the coherent
evolution in the TC ladder shown in Fig.\,\ref{fig:levels}, reproduces the rich features of the
measurements quantitatively.

Modifying the detuning, we can adjust the system to exhibit only one exciton in resonance with the
cavity, while the other excitons are significantly detuned, so that the dynamics resembles that of
a simpler single-exciton cavity system. This is achieved at $\delta=-133\,\mu$eV (T=13.5\,K), for
which X1 is in resonance with C within $5\,\mu$eV, whilst X2 and X3 are detuned by $135\,\mu$eV and
$253\,\mu$eV, respectively.  The measured and predicted $|\tilde{P}(\omega,\tau)|^2$ and
$|P(t,\tau)|^2$ are presented in Fig.\,\ref{fig:FWMDynamics}f,h and Fig.\,\ref{fig:FWMDynamics}g,i,
respectively, while the time-integrated FWM is displayed in Fig.\,\ref{fig:FWMDynamics}j. We
observe a beat versus delay $\tau$ with a period of about 50\,ps, corresponding to a polaritonic
splitting of $83\,\mu$eV, somewhat larger than the calculated splitting of $60\,\mu$eV, which is
slightly below $2g_1$ due to the finite damping. The faster than expected beat period is due to
remaining influence of the two additional excitons, as shown by the agreement of the predicted
dynamics including all excitons (see solid line) with the measurements.

\begin{figure}
\iffigs\ifpdf\includegraphics*[width=\columnwidth]{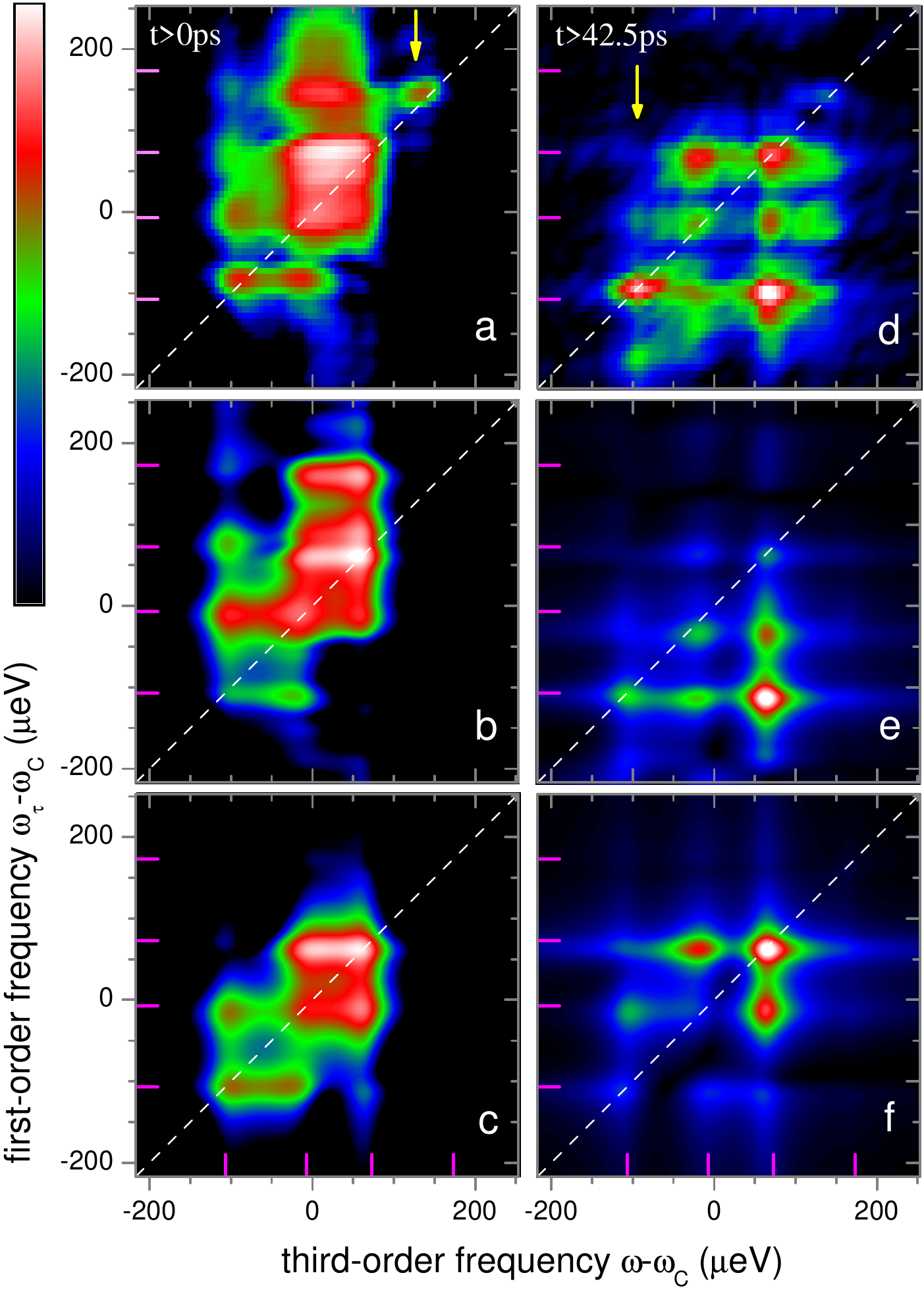}\fi\fi \caption{ Two-dimensional FWM at
T=19\,K. Power $|\bar{P}(\omega,\omega_\tau)|^2$, measured (a) and predicted with (b) and without
(c) phase correction. Logarithmic colour scale from 0.09 (black) to 1 arbitrary units. Yellow
arrows indicate the $\omega$ of the phase correction. Magenta ticks indicate the polariton
frequencies of the first rung $\lambda_{1,k}$ (see SM). (d,e,f) as (a,b,c), but showing the
post-selected $|\bar{P}_{\rm s}(\omega,\omega_\tau,42.5\,{\rm ps})|^2$ on a linear colour scale
from 0.01 to 0.24 arbitrary units. \label{fig:2DFWM}}
\end{figure}

The agreement between the measured FWM dynamics and independently predicted FWM in the framework of
the cavity-mediated coupling model for different detuning is revealing the coherent wiring up of
the three Xs via the cavity mode in the studied microresonator. A definite display of the coherent
coupling is afforded by the two-dimensional (2D) frequency domain representation of the FWM, in
which coherent coupling is observed as off-diagonal signals \cite{MukamelARPC00, ColliniNature10,
ScholesNatureChem11}. We retrieve the 2D FWM \cite{LangbeinOL06} by Fourier-transforming
$\tilde{P}(\omega,\tau)$ from the delay time $\tau$ into the conjugated frequency $\omega_\tau$
yielding $\bar{P}(\omega,\omega_\tau)$. In this transformation we use only positive delays
$\tau>0$, such that $\omega_\tau$ represents the frequency of the first-order polarization created
by $E_1$. The resulting two-dimensional FWM diagrams $\bar{P}(\omega,\omega_\tau)$ are presented in
Fig.\,\ref{fig:2DFWM}a,d. In this representation, coupled resonances manifest themselves by
corresponding off-diagonal components, showing that resonances excited by the first pulse in
first-order ($\omega_\tau$ axis) coherently couple to a different resonances emitting in
third-order ($\omega$ axis). To enable the Fourier transformation we set the relative phase of data
at different delays $\tau$ using a phase correction \cite{KasprzakNaturePhot11} at a given $\omega$
(see yellow arrows) as discussed in the SM. The effect of the phase-correction is illustrated in
the difference between the predictions with and without phase correction given in
Fig.\,\ref{fig:2DFWM}b,c, respectively. The phase correction leads to a weighting in the
first-order frequency, but does not qualitatively change the off-diagonal structure.

Due to the contributions of the 28 second-rung transitions, the features in 2D FWM are broadened.
Post-selecting the FWM signal emitted after a survival time $t_{\rm s}$ after its creation by the
second pulse at $t=0$, resulting in the 2D FWM  $\bar{P}_{\rm s}(\omega,\omega_\tau,t_{\rm s})$, we
can suppress the fast decaying components \cite{KasprzakNatureMat10}. The measured and predicted
post-selected  $\bar{P}_{\rm s}(\omega,\omega_\tau,42.5\,{\rm ps})$ (see
Fig.\,\ref{fig:2DFWM}d,e,f) are dominated by the first-rung transitions and display a clear
separation between the multiple off-diagonals at the polariton frequencies $\lambda_{1,k}$ (see
SM), showing cavity-mediated mutual coherent coupling between Xs. The corresponding data for
$T=13.5\,$K ($\delta=-133\,\mu$eV) given in the SM, Fig.\,S6, shows the coherent coupling between
two polaritonic modes dominated by the cavity and X1. A detailed analysis of the strength of the
off-diagonal peaks and their relation to the coherent coupling strength will be presented in a
forthcoming work.

The observed coherent coupling between three qubits via the cavity mode demonstrates that an
optical microcavity can act as a coupling bus for excitonic qubits. The coherent interaction can be
controlled by non-local tuning of the cavity mode or the exciton energy, or by switching the
excitonic qubit between bright and dark exciton states \cite{PoemNaturePhys10}. The coupling via
the cavity mode has the prospect to be spatially extended in coupled cavity structures, for example
in photonic crystals \cite{EnglundOE07}. The small size of the quantum dots with respect to the
cavity enables coupling of many more than three qubits, enhancing the prospects of quantum
information processing on a chip with a photonic interface.\\

{\bf Methods}\\

{\bf Samples:} The fabrication process of high-Q quantum dot micropillar structures starts with the
growth of a planar microcavity on an undoped GaAs substrate by molecular beam epitaxy. A GaAs
$\lambda$-cavity with a single layer of self-assembled In$_{0.4}$Ga$_{0.6}$As QDs with a ground
state exciton emission wavelength around 930\,nm is sandwiched between two distributed Bragg
reflectors (DBR) providing a vertical photon confinement. The upper (lower) DBR consists of 26 (30)
mirror pairs, which are made of $\lambda/4$ layers of GaAs and AlAs. For lateral photon confinement
micropillars with a nominally circular cross-section and diameters between $1.6\,\mu$m and
$2\,\mu$m were structured by means of high resolution electron beam lithography and electron
cyclotron etching. The fundamental cavity mode in the investigated structure has a Q-factor of
30000, and a corresponding photon lifetime of about 10\,ps within its mode volume of approximately $0.3\,\mu{\rm m}^3$.\\

{\bf FWM experiment:} We employ heterodyne spectral interferometry\cite{LangbeinOL06}, using a pair
of pulse trains, $E_1$ and $E_2$ at a repetition rate of 76\,MHz and 1\,ps pulse-duration which are
frequency up-shifted by $(\Omega_1/2\pi, \Omega_2/2\pi)=(79, 80.7)$\,MHz by acousto-optic devices.
The spatial mode of $E_1$ and $E_2$ is matched to the cavity mode on the top facet of the
micropillar by a microscope objective mounted inside a helium bath cryostat. The pulses are
resonant to the cavity mode and are injecting into the intra-cavity field in average 0.25 (0.75)
photons per pulse of $E_1$ ($E_2$), respectively, see Fig.\,S7 in SM. The emitted light is
interfered with a frequency unshifted reference field and the FWM signal is selected at the
heterodyne beat note $2\Omega_2-\Omega_1$ and spectrally resolved. The complex FWM response
$P(t,\tau)\propto E_1^*E_2^2$ is retrieved by
spectral interferometry \cite{LangbeinOL06,LangbeinJPCM07}.\\

{\bf Theory:} We use a standard approach \cite{ReitzensteinOL06,LauchtPRB10,LaussyPRB11}, with the
Hamiltonian of a single photonic mode coupled to three two-level systems with dissipation of all
components taken into account by a Lindblad super-operator. To calculate the FWM polarization, we
follow our earlier rigorous approach \cite{MuljarovPSS06,KasprzakNatureMat10} to the optical
response of a system on a sequence of optical pulses. For the present system excited by two
ultrashort pulses it has the analytic form
\be P_{\rm FWM}(t,\tau)= \sum_{jk} c_{jk} e^{i\tilde{\omega}_j t} e^{i\lambda_k \tau} \ee
in which $\lambda_k=\lambda^{\ast}_{1k}$ ($\lambda_k=\lambda_{2k}$) are the complex polariton
frequencies of the first (second) rung, for positive (negative)  delay times $\tau$, respectively,
while $\tilde{\omega}_j$ are the complex frequencies of all possible transitions between the ground
state and the first rung and between the first and second rung, all contributing to the real-time
dynamics.

The expansion coefficients $c_{jk}$ are calculated exactly reducing the full master equation for
the density matrix to a finite matrix problem. Further details are available in the SM.\\

\begin{acknowledgments}
F.A. acknowledges support by the Deutscher Akademischer Austauschdienst under project number
50743470. The work was in part supported by the Deutsche Forschungsgemeinschaft through the
research  group `Quantum Optics in Semiconductor Nanostructures' and the State of Bavaria. K.S.
acknowledges support by the Royal Thai Government.\\
\end{acknowledgments}

{\bf Author contributions:} The work was initiated by SR and WL. The sample was grown and processed
by SR, CK, CS, MS, SH and AF, and characterized by FA and SR. Measurements were performed by FA,
JK, and WL, and analyzed by FA, JK, WL and EM. The theory was developed and compared with the
experiment by KS, EM and WL. The manuscript was written by FA, JK, SR, EM, and WL.

The authors declare no competing financial interests.

%\bibliography{paper,langsrv}
\bibliography{MQD_Manuscript}

\end{document}

\hfill {Online supplementary material}

%\vskip5mm
\bigskip

%\preprint{}

\title{Microcavity controlled coupling of excitonic qubits}



\author{F. Albert}
\affiliation{Technische Physik, Physikalisches Institut, Universit\"at W\"urzburg and Wilhelm Conrad R\"ontgen Research
Center for Complex Material Systems, Am Hubland, D-97074 W\"urzburg, Germany}
\author{K. Sivalertporn}
\affiliation{Cardiff University School of Physics and Astronomy, The Parade, Cardiff CF24 3AA, United Kingdom}
\author{J. Kasprzak}
\affiliation{Institut N\'eel, CNRS et Universit\'e Joseph Fourier, BP 166, F-38042 Grenoble Cedex 9, France}
\author{M Strau\ss}
\author{C. Schneider}
\author{S. H\"ofling}
\author{M. Kamp}
\author{A. Forchel}
\affiliation{Technische Physik, Physikalisches Institut, Universit\"at W\"urzburg and Wilhelm Conrad R\"ontgen Research
Center for Complex Material Systems, Am Hubland, D-97074 W\"urzburg, Germany}
\author{S. Reitzenstein}
\altaffiliation[Present Address: ] {Institut f\"ur Festk\"orperphysik, Technische Universit\"at Berlin,
Hardenbergstrasse 36, 10623 Berlin, Germany}\affiliation{Technische Physik, Physikalisches Institut, Universit\"at
W\"urzburg and Wilhelm Conrad R\"ontgen Research Center for Complex Material Systems, Am Hubland, D-97074 W\"urzburg,
Germany}
\author{E. A. Muljarov}
\author{W. Langbein}
\affiliation{Cardiff University School of Physics and Astronomy, The Parade, Cardiff CF24 3AA, United Kingdom}
\email{langbeinww@cardiff.ac.uk}


\date{\today}

%\pacs{}
\keywords{}

\maketitle

\section{Four-wave mixing response of $N$ quantum dots coupled to the cavity}

The Tavis-Cummings (TC) model of $N$ quantum dots (QDs) coupled to a single photonic mode in a
microcavity is described by the Hamiltonian
 \be
H= \wc a\dag a +\sum_{n=1}^N \biggl[\wxn |n\rangle\langle n|+g_n \Bigl( a\dag |0\rangle\langle n| +
a |n\rangle\langle0|\Bigr)\biggr]\,,
 \ee
in which $\wc$ is the frequency of the cavity mode, $\wxn$ is exciton transition energy of the
$n$-th QD, and $g_n$ is the coupling strength between the $n$-th QD excition and the cavity mode.
Furthermore $a\dag$ ($a$) is the cavity photon creation (destruction) operator, $|0\rangle$ is the
ground state (GS) of the exciton system, and $|n\rangle$ denotes the state with one exciton in the
$n$-th QD and all other QDs empty. We use $\hbar=1$ for simplicity of notations. We assume that all
QD excitons are close to resonance with the cavity mode, and neglect any excited or multi-excitonic
states within the same QD, as they are generally off resonance. However, since we consider QDs
which are electronically uncoupled, multi-excitonic states in which all excitons are in different
QDs stay close to resonance with the cavity mode, and significantly contribute to the four-wave
mixing (FWM) dynamics.

The system of the $N$ QD excitons coupled to the cavity is excited by an external electric field
consisting of two ultrashort pulses separated by the delay time $\tau$ and having complex pulse
areas $\mu E_1$ and $\mu E_2$, where $\mu$ is the effective dipole moment of the cavity mode. The
measured FWM polarization is a third-order signal proportional to $E_1^\ast E_2^2$. In the
calculation, the effect of the pulses is taken into account as follows. The first pulse $E_1$
contributes in first-order, changing the density matrix according to
 \be
\r^{(+)}=-i\mu E_1^\ast[a,\r^{(-)}]\,,
\label{first}
 \ee
where $\r^{(-)}$ ($\r^{(+)}$) is the density matrix before (after) the pulse. The second pulse $E_2$ contributing in
second order results in the following change of the density matrix:
 \be
\r^{(+)} = \frac{(-i)^2}{2}\mu^2 E_2^2[a\dag,[a\dag,\r^{(-)}]]\,.
\label{second}
 \ee
Between and after the pulses the time evolution of the density matrix $\r(t)$ is described by the
master equation
 \be
i\frac{d\r}{dt} = \L \r\,,
 \label{master}
 \ee
where $\hat{L}$ is the Lindblad super-operator
of the exciton-cavity system,
 \be
\L \r = [H,\r] - i \gc \bigl(a\dag a \r +\r a\dag a - 2 a\r
a\dag\bigr)  + \sum_{n=1}^N \gxn \biggl[ |n\rangle\langle n|\rho + \rho|n\rangle\langle n| - 2|0\rangle\langle n|\rho|n\rangle\langle0|\biggr]\,,
\label{Lin}
 \ee
with $\gc$ and $\gxn$ being, respectively, the cavity damping and the exciton dephasing, which are
assumed to be lifetime limited. Equation (\ref{master}) has the following formal solution
 \be
\r(t)=e^{-i\L (t-t')}\r(t')\,,
 \label{fme}
 \ee
and thus the FWM polarization takes the form \be P(t,\tau)={\rm Tr}\{\r\, a\}= \frac{(-i)^3}{2} \mu^3 E_1^\ast
E_2^2\cdot {\rm Tr}\Bigl\{R(t,\tau) a\Bigr\} \label{PFWM} \ee  where the third-order component of the density matrix is
proportional to
 \be
R(t,\tau) = e^{-i\L t} \Bigl[ a\dag , \bigl[a\dag, e^{-i\L
\tau }[a, \r(-\infty)]\bigr]\Bigr]\,,
 \label{r3plus}
 \ee
for positive delay between the pulses ($\tau>0$), and to
 \be
R(t,\tau) =  e^{-i\L t} \Bigl[ a , e^{i\L \tau } \bigl[a\dag,
[a\dag, \r(-\infty)]\bigr]\Bigr]\,,
 \label{r3minus}
 \ee
for negative delay ($\tau<0$). Here $\r(-\infty)$ is the density matrix before the excitation.

We solve equations (\ref{first})--(\ref{master}) {\em exactly} by expanding all operators into a
set of uncoupled exciton-photon states $|{\it
i}\rangle\equiv|\nxone,\nxtwo,\ldots,\nxnn;\nc\rangle$ ($i={\it 0,\,1,\,2}\dots$), where
$\nxn=0,\,1$ and $\nc=0,\,1,\,2\dots$ are, respectively, the exciton and photon occupation numbers.
We assume that the system is initially in its ground state, with the density matrix
$\rho(-\infty)=|{\it 0}\rangle\langle {\it 0}|$. This is justified by the excitations of the system
having an energy about 3 orders of magnitude larger than the thermal energy and coherence times
shorter than the period of the repetitive excitation (13\,ns) used in this work. The resulting FWM
dynamics includes only transitions between the GS, the first rung, and the second rung of the TC
ladder. With $N$ quantum dots coupled to the cavity, there are $N_1=1+N$ states in the first rung
and $N_2=1+N(N+1)/2$ states in the second rung, so that the expansion of density matrix has the
form
 \be
 \r(t)=\sum_{i,j=0}^{N_1+N_2}\r_{ij}(t)\,|i\rangle \langle j|\,,
 \ee
with $1+N_1+N_2$ basis states. A similar expansion is used for the photon creation (annihilation)
operator contributing to Eqs.\,(\ref{first}) and (\ref{second}). To solve the master equation
(\ref{master}) after the pulsed excitation, a bigger, $M\times M$ matrix of the super-operator
$\hat{L}$ is introduced, with $M=N_1(1+N_2)$ being the total number of all possible transitions,
between the GS and the first rung ($N_1$), and between the first and second rungs ($N_1N_2$). Only
such transitions contribute to the FWM polarization. The solution to Eq.\,(\ref{master}) then has
the matrix form
 \be
\r(t)=e^{-i\L t}\r(0)=\hat{U}
e^{-i\hat{\Omega}t}\hat{V}\r(0)\,,
 \label{fme1}
 \ee
in which the matrix $\L$ is diagonalised as
 \be
\L =\hat{U} \hat{\Omega} \hat{V}\,,\ \ \ \ \hat{U} \hat{V} =
\hat{1}\,,
 \ee
with $\hat{U}$ and $\hat{V}$ being matrices of right and left eigenvectors and $\hat{\Omega}$ a
diagonal matrix of complex eigenfrequencies $\tilde{\omega}_j$ of all possible transitions
involved. The density matrix then becomes a finite superposition of exponentials, \be
\r(t)=\sum_{j=1}^M r_j e^{-i\tilde{\omega}_j t} \ee with the frequencies $\tilde{\omega}_j$ and
matrices $r_j$ calculated exactly. In particular, \be \tilde{\omega}_j=\lambda_{1,k}\ \ \ {\rm and
} \ \ \ \tilde{\omega}_{j}=\lambda_{2,k}-\lambda_{1,m}^\ast \ee for the GS\,--\,first rung and
first rung\,--\,second rung transitions, respectively, where $\lambda_{1,k}$ ($\lambda_{2,k}$) are
the complex energy levels of the first (second) rung, calculated by diagonalizing of the complex
symmetric matrix of a non-Hermitian Hamiltonian (see below). The time evolution between pulses is
given by a smaller matrix $\hat{L}$ as only transitions from the GS to the first (second) rung
participate in the delay dynamics, for positive (negative) delay times. The FWM polarization then
takes the following explicit form: \be P(t,\tau)=\sum_{j=1}^M e^{i\tilde{\omega}_j t}\times\left\{
\begin{array}{l l} \displaystyle
\sum_{k=1}^{N_1} a_{jk} e^{i\lambda_{1,k}^\ast \tau} & \tau>0 \\
\displaystyle
\sum_{k=1}^{N_2} b_{jk} e^{i\lambda_{2,k} \tau} & \tau<0 \\
\end{array}
\right. \label{Pttau} \ee With the analytic form Eq.\,(\ref{Pttau}) at hand, it is straightforward to calculate the
Fourier transform (FT) of the FWM polarization. In particular, for positive delays, its two-dimensional FT has the form
\be P(\omega,\omega_\tau)=\sum_{j=1}^M \sum_{k=1}^{N_1}
\frac{a_{jk}}{(\omega-\tilde{\omega}_j+i\gs)(\omega_\tau+\lambda_{1,k}^\ast)}\,, \ee where a Lorentzian spectrometer
resolution $\gs$ (half width at half maximum) is included.

Although our analytical approach is general and can be used for any arbitrary $N$ producing a reasonable size of the basis (the size of the matrix $\hat{L}$ scales as $N^3$), in the present calculation we concentrate on a system of $N=3$ QDs coupled to the cavity. Then the basis reduces to the following {\it twelve} states which include $N_1=4$ states of first rung and $N_2=7$ states of the second rung:
 \bea
\texttt{\rm Ground\ state} \ &&|{\it 0}\rangle = |0,0,0;0\rangle;\nonumber\\
&&\nonumber\\
\texttt{\rm First\ rung} \  &&|{\it 1}\rangle = |0,0,0;1\rangle\,,\ \ |{\it 2}\rangle = |1,0,0;0\rangle\,,\ \ \ \,
|{\it 3}\rangle = |0,1,0;0\rangle\,,\ \ |{\it 4}\rangle = |0,0,1;0\rangle;\nonumber\\
&&\nonumber\\
\texttt{\rm Second\ rung} \ &&|{\it 5}\rangle = |0,0,0;2\rangle\,,\ \ |{\it 6}\rangle = |1,0,0;1\rangle\,,\ \ \ \, |{\it 7}\rangle = |0,1,0;1\rangle\,,\ \ |{\it 8}\rangle = |0,0,1;1\rangle\,,\nonumber\\
&&|{\it 9}\rangle = |1,1,0;0\rangle\,,\ \ |{\it 10}\rangle = |1,0,1;0\rangle\,,\ \
|{\it 11}\rangle = |0,1,1;0\rangle\,.
\label{states}
 \eea
Both the density matrix and the photon creation operator are 12$\times$12 matrices, while the
super-operator $\hat{L}$ after the excitation by both pulses is represented by a $32\times32$
matrix which is due to $M=32$ transitions: 4 transitions between the GS and the first rung and
$4\times7=28$ transitions between the first and second rungs.

The energy levels $\lambda_{n,k}$ of the TC ladder are calculated by diagonalizing the effective
non-Hermitian Hamiltonian:
 \be
\tilde{H} = \left( \begin{array}{c c c c} \tilde{H_1} & 0 & 0&\dots\\
0&\tilde{H_2} & 0 &\dots\\
0&0&\tilde{H_3} &\dots\\
 \vdots & \vdots& \vdots &\ddots\end{array}\right)\,.
 \label{Heff}
 \ee
The first two blocks on the main diagonal of $\tilde{H}$ refer to the first and second rungs, respectively:
 \be
\tilde{H_1} = \left( \begin{array}{c c c c}
\twc & g_1 & g_2 & g_3 \\
g_1 & \twxone & 0 & 0 \\
g_2 & 0 & \twxtwo & 0 \\
g_3 & 0 & 0 & \twxthree \\
\end{array}\right)\,,
 \label{Heff1}
 \ee
 %\begin{widetext}
 \be
\tilde{H_2} = \left( \begin{array}{c c c c c c c}
2 \twc & \sqrt{2}g_1 & \sqrt{2}g_2 & \sqrt{2}g_3 & 0 & 0 & 0 \\
\sqrt{2}g_1 & \twc+\twxone & 0 & 0 & g_2 & g_3 & 0 \\
\sqrt{2}g_2 & 0 & \twc+\twxtwo & 0 & g_1 & 0 & g_3 \\
\sqrt{2}g_3 & 0 & 0 & \twc+\twxthree & 0 & g_1 & g_2 \\
0 & g_2 & g_1 & 0 & \twxone+\twxtwo & 0 & 0 \\
0 & g_3 & 0 & g_1 & 0 & \twxone+\twxthree & 0 \\
0 & 0 & g_3 & g_2 & 0 & 0 & \twxtwo+\twxthree \\
\end{array}\right)\,,
 \label{Heff2}
\ee
%\end{widetext}
where $\twc=\wc-i\gc$ and $\twxn=\wxn-i\gxn$.\\

\section{Photoluminescence characterization}

\begin{figure}
\begin{center}
\iffigs\ifpdf \includegraphics*[width=0.7\columnwidth]{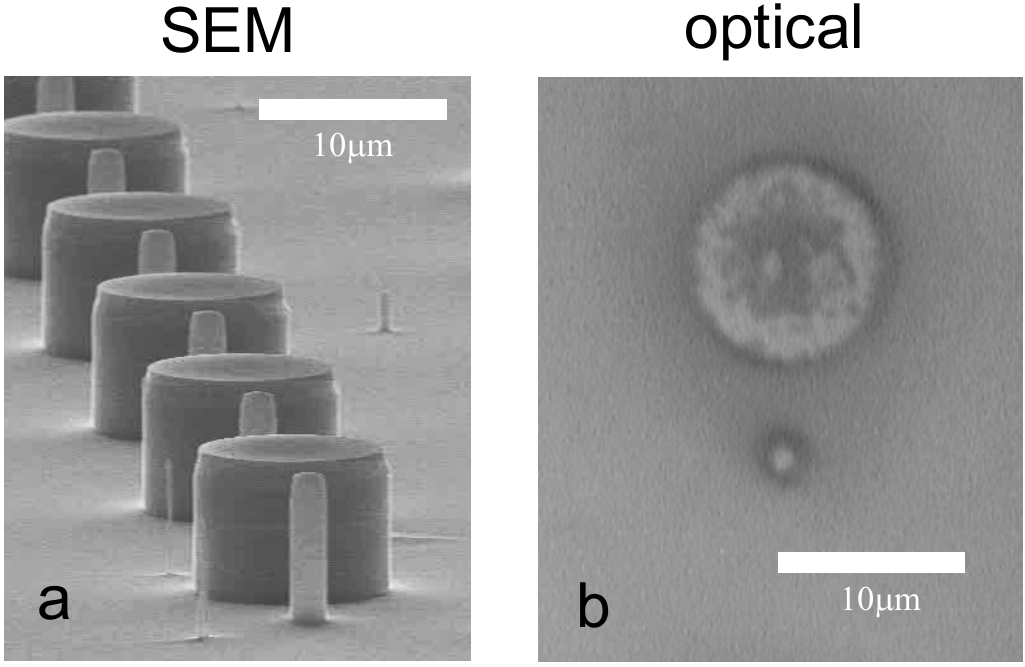}\fi\fi \caption{Images of the
investigated sample. a) Scanning electron microscopy image of a row of micropillar - reference
pillar pairs. b) White light reflection image taken in the cryostat at 10\,K using the same optical
setup as in the confocal PL and HSI experiments. Scale bars are given. \label{fig:SM_Sample}}
\end{center}
\end{figure}

The investigated sample consists of fields of 6 by 6 micropillars of nominally equal diameter, with
$50\,\mu$m separation in $x$ and $y$ direction corresponding to the $[110]$ and $[1\bar{1}0]$
crystallographic directions of the GaAs substrate, respectively. Each micropillar was accompanied
by a reference pillar of $10\,\mu$m diameter at $10\,\mu$m distance which was used to reflect the
reference beam in the heterodyne spectral interferometry (HSI) experiment. A scanning electron
microscopy image of a row is shown in Fig.\,\ref{fig:SM_Sample}a.

The micropillars have diameters of 1.5, 1.6, 1.7, 1.8 and $1.9\,\mu$m, and for each diameter 10
equal fields provided 360 nominally equal micropillars. The sample was mounted on an $x-y$ stage in
a helium bath cryostat, together with a microscope objective of 0.85 numerical aperture which was
adjustable in all three dimensions by a home-build stage using piezoelectric bender actuators. A
reflection image of a micropillar taken with this arrangement is shown in
Fig.\,\ref{fig:SM_Sample}b. To identify micropillars in the strong coupling regime, we used
confocal photoluminescence ($\mu$PL) excited by a laser of 532\,nm wavelength, focussed to a spot
of about $0.5\,\mu$m FWHM on the top face of the micropillar. Typically 10\% of the micropillars
with $1.5-1.7\,\mu$m diameter showed strong coupling. The micropillar presented in this work was
selected from about 200 micropillars for the strongest coupling of multiple quantum dots. The
photoluminescence excitation power at the micropillar was about 100\,nW. At lower intensities the
PL spectrum remained essentially unchanged, while at higher intensities line broadening and
reduction of the Rabi-splitting was observed.

\begin{figure}
\begin{center}
\iffigs\ifpdf\includegraphics*[width=0.6\columnwidth]{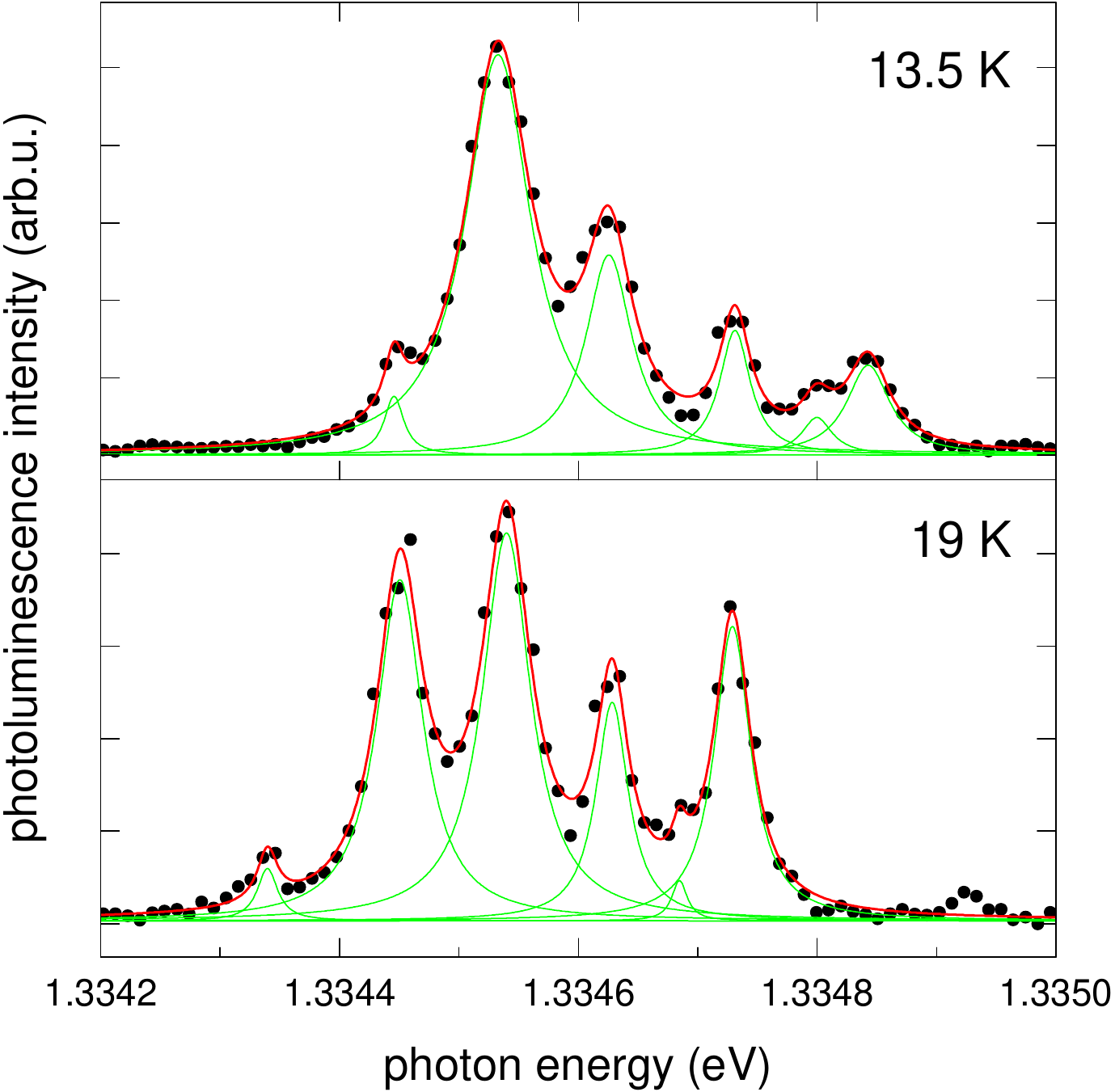}\fi\fi \caption{Photoluminescence
spectra from the investigated micropillar at a temperature of 13.5\,K (top) and 19\,K (bottom), and
fits by multiple Lorentzian lines (solid lines). \label{fig:SM_PLFit} }
\end{center}
\end{figure}

To characterize the quantum dot - cavity coupling, $\mu$PL spectra were measured as function of the
sample temperature from 7\,K to 35\,K. The PL spectrum at each temperature was fitted by a sum of
Lorentzian lines, as exemplified in Fig.\,\ref{fig:SM_PLFit}. The resulting line positions as
function of temperature were fitted with the first rung transition energies $\lambda_{1,k}$ given
by \Eq{Heff1}. In the fit, we used an explicit functional dependence for the temperature-tuning of
the energies of quantum dots and cavity, from a model of the band-gap shift of semiconductors
\cite{PasslerJAP99}.

\be\wxn(T)=\wxn(0)+F(T) ,\quad \wc(T)=\wc(0)+\eta F(T) \label{eqn:wTdep}\ee with \be F(T)
=-\frac{\alpha\theta}{2}\left(\coth\left(\frac{\theta}{2T}\right) - 1\right)\ee

The parameters describing the temperature-dependence were determined from a fit to the emission of
quantum dots detuned from the cavity by more than 1 meV, yielding $\alpha = (60.9\pm 0.6)\,\mu$eV/K
and  $\theta = (58.9\pm 1.0)\,K$. The quoted errors represent the statistical variation of the
results between different fitted quantum dots.

From the fit of $\lambda_{1,k}$ to the measured line positions and widths, we deduce the low
temperature energies $\wxn(0), \wc(0)$, the linewidths $\gxn,\gc$ and the coupling strengths $g_n$,
as shown in table \ref{tab:systemfit}. Additionally, the scaling factor $\eta=0.227\pm 0.003$, for
the cavity shift was determined, showing that the cavity shift is about 4 times less than the
quantum dot exciton shift.

\begin{table}
\begin{center}
\begin{tabular}{c|c|c}
\hline
parameter & value & unit\\
\hline
$\wxone(0)$ & 1.3346106 & eV\\
$\wxtwo(0)$ & 1.3347412 & eV\\
$\wxthree(0)$ & 1.3348584 & eV\\
$\wc(0)$ & 1.3345732 & eV\\
\hline
$g_1$ &  43 & $\mu$eV\\
$g_2$ &  40 & $\mu$eV\\
$g_3$ &  31.5 & $\mu$eV\\
\hline
$\gxone$ &  18 & $\mu$eV\\
$\gxtwo$ &  11.5 & $\mu$eV\\
$\gxthree$ &  16 & $\mu$eV\\
$\gc$ &  36.5 & $\mu$eV\\
\hline
\end{tabular}
\end{center} \caption{Parameters of the three-exciton cavity system deduced from a fit to the temperature-dependent photoluminescence spectra.} \label{tab:systemfit}
\end{table}

For simplicity, we assume here that the linewidths and the coupling strengths are independent of
temperature, and that the linewidths are Lorentzian, i.e. homogeneously broadened. A spectrometer
resolution of $\gs=4\,\mu$eV (HWHM) was subtracted from the fitted linewidths.

The intrinsic homogeneous linewidth of the quantum dots\cite{BorriPRB05} is expected to be limited
by radiative decay and phonon scattering. The radiative decay time of the quantum dots in bulk GaAs
is about 400\,ps, corresponding to $1\,\mu$eV HWHM. In micropillars, this decay is not
significantly modified due to the presence of leaky modes which have a similar local density of
states as the bulk modes \cite{BayerPRL01}.  The phonon-scattering increases the zero-phonon
linewidth for quantum dots with 140\,meV confinement energy \cite{BorriPRB05} by about $0.5\,\mu$eV
at 20\,K and $2\,\mu$eV at 30\,K.

The measured linewidths of different uncoupled quantum dots are between 10 and $30\,\mu$eV, and
vary from dot to dot, both in terms of low-temperature linewidth and dependence on temperature,
which we attribute to spectral diffusion. Such a broadening mechanism is presently not taken into
account in the FWM modeling, and might explain some of the remaining differences between the
experimental data and the modeling, specifically for negative delay due to the expected photon echo
formation. We plan to extend the modeling to include also inhomogeneous broadening and pure
dephasing of the quantum dot excitons \cite{PattonPRB06, SeidlPRB05} in future work.

The coupling strength is expected to decrease with temperature proportional to the square root of
the zero-phonon line weight\cite{BorriPRB05}, resulting in a reduction of about 5\% from 8\,K to
30\,K. We have neglected this dependence. The resulting fit (see Fig.\,1) reproduces the energy
positions and linewidths to within the measurement error of about $5\,\mu$eV. The above
approximations are therefore resonable for the data at hand. The linewidths of the strongly coupled
modes are actually dominated by the cavity linewidth, which is given by the photon lifetime and
thus Lorentzian. This reduces the importance of the broadening mechanism in the quantum dot
excitons.

\section{Polariton states}

\begin{figure}[h]
\iffigs\ifpdf \includegraphics*[width=0.45\columnwidth]{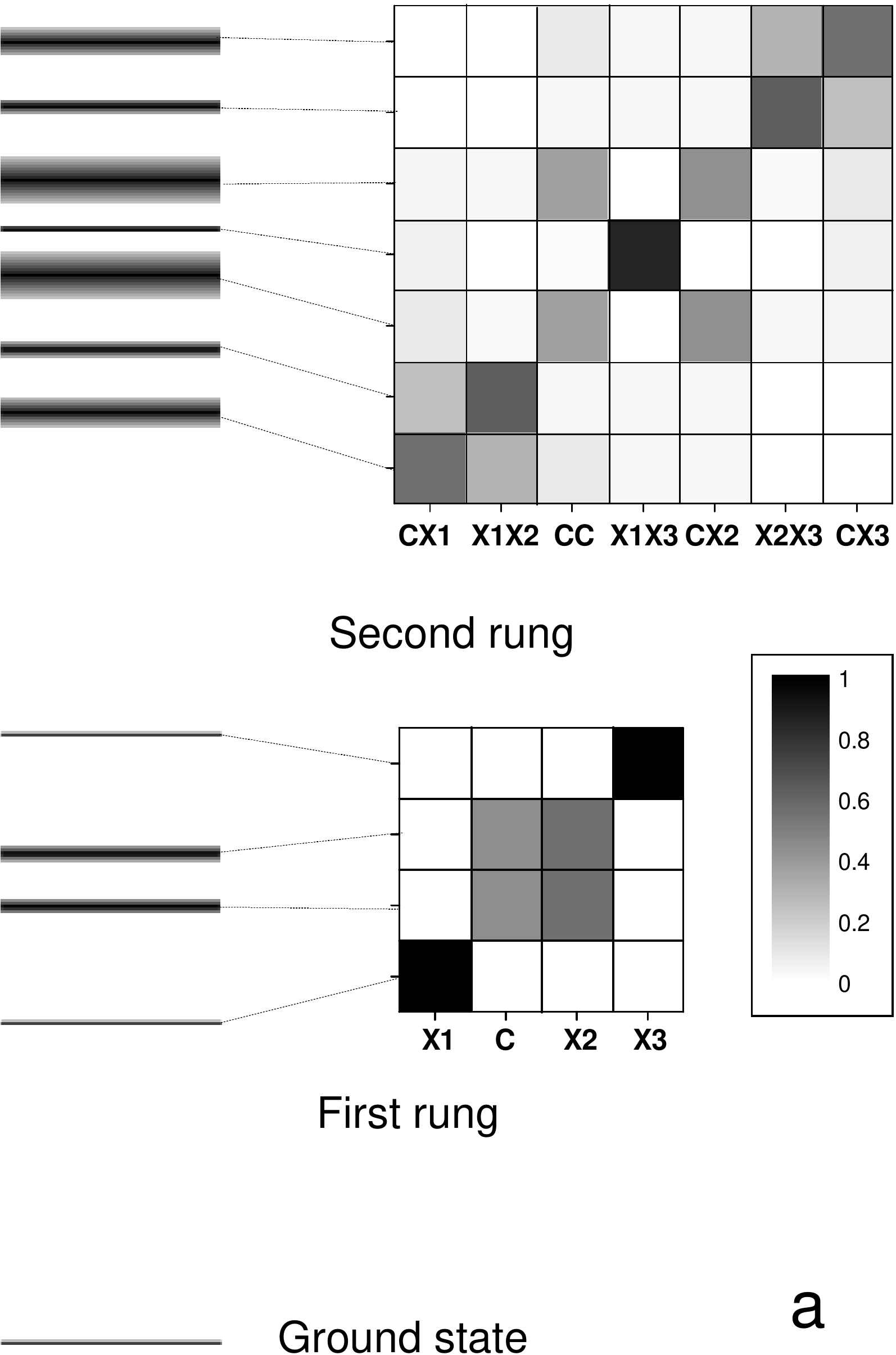} \hfill
\includegraphics*[width=0.45\columnwidth]{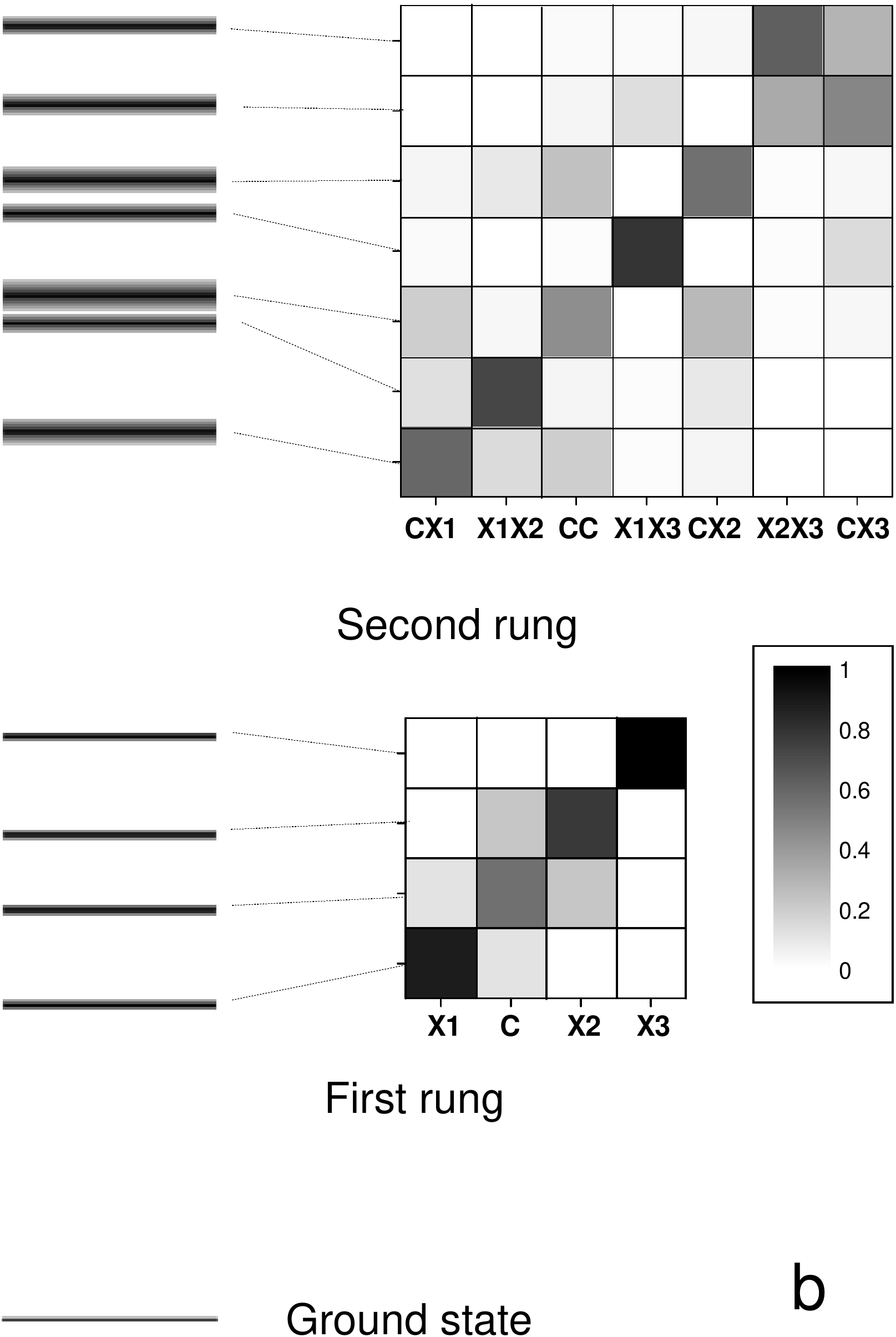} \fi\fi
\caption{Energies and linewidths of the polariton states of the first and second rungs and their
probability distribution (grey scale matrices) in the basis of the uncoupled states
Eq.\,(\ref{states}) of the exciton-cavity system, calculated for (a) model symmetric structure and
(b) realistic asymmetric structure with $\delta=-29\,\mu$eV (corresponding to $T=19$\,K). The
parameters of the system used for the symmetric structure are $g_n=40\,\mu$eV, $\gxn=0$,
$\gc=60\,\mu$eV, $\wxtwo=\wc$, and $\wxtwo-\wxone=\wxthree-\wxtwo=130\,\mu$eV (corresponding to
zero detuning). The parameters of the realistic structure are given in the main text and
table\,\ref{tab:systemfit}. Here, the similar contributions of the cavity mode to all states
results in similar linewidths of the levels of a given rung. \label{fig:SM_Levels} }
\end{figure}

\begin{figure}
\begin{center}
\iffigs\ifpdf \includegraphics*[width=0.6\columnwidth]{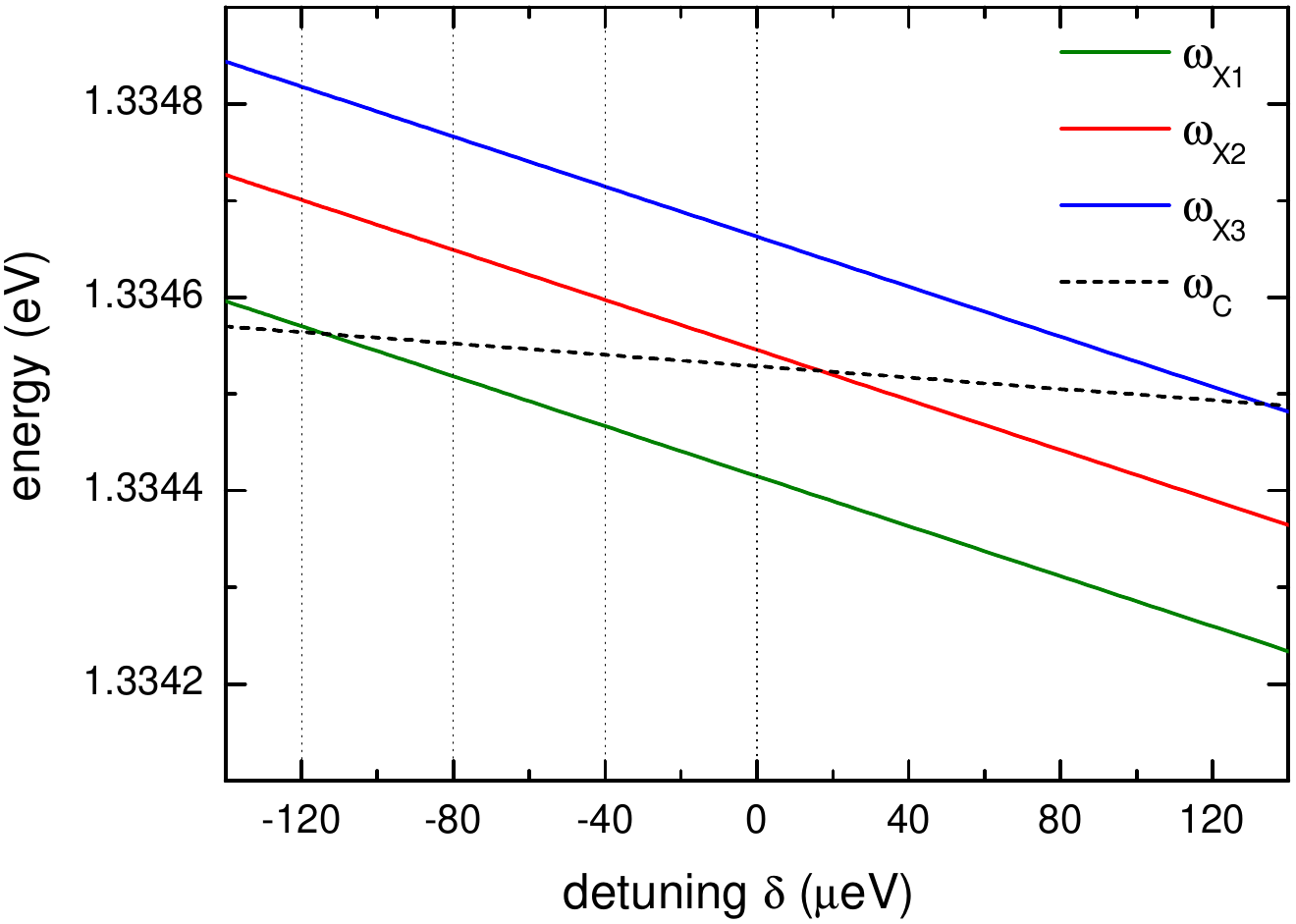} \fi\fi
\end{center}
\caption{Energies of the uncoupled QD-exciton and cavity modes as function of the detuning
$\delta=\wc-\left(\sum_{n=1}^3
  g_n \wxn\right)/ \left(\sum_{n=1}^3 g_n\right)$. Dotted vertical lines correspond to the values of the detuning used in Fig.\,\ref{fig:SM_2DFWM_Detuning}.
\label{fig:SM_Detuning} }
\end{figure}

The polariton states in the micropillar calculated for a symmetric model structure and realistic parameters of the
investigated sample are shown in Fig.\,\ref{fig:SM_Levels}. The detuning dependence of the cavity and exciton levels
are shown in Fig.\,\ref{fig:SM_Detuning}.

\begin{figure}%[t]
\begin{center}
\iffigs\ifpdf\includegraphics*[width=0.9\columnwidth]{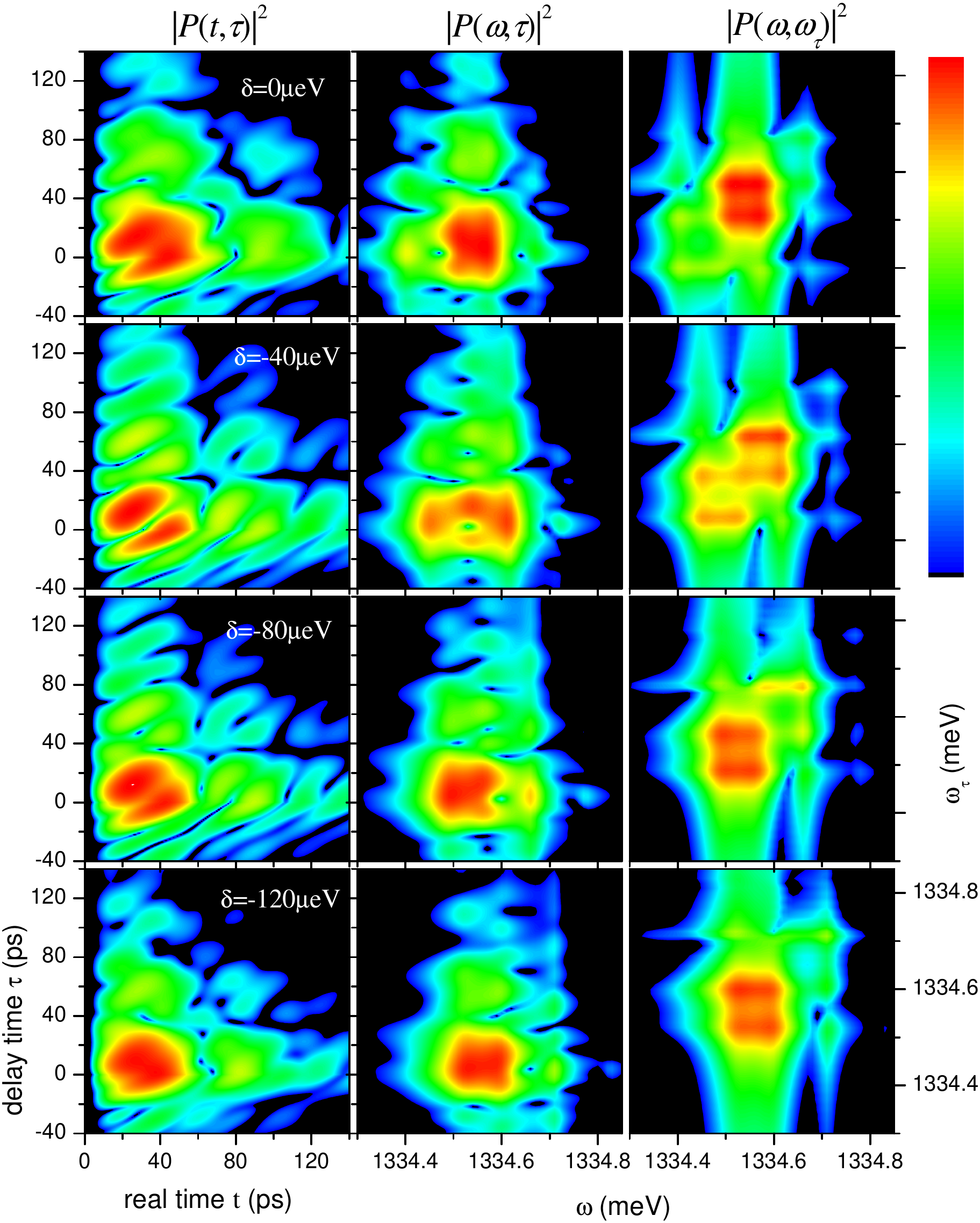}\fi\fi
\end{center}
\caption{Calculated FWM versus detuning using the parameters listed in table \ref{tab:systemfit}.
FWM intensities $|P(t,\tau)|^2$ (left column), $|\tilde{P}(\omega,\tau)|^2$ (middle column), and
$|\bar{P}(\omega,\omega_\tau)|^2$ (right column), for the detunings $\delta=(0,-40,-80, -120)\mu$eV
from top to bottom. Logarithmic colour scale as shown over 4 orders of magnitude.
\label{fig:SM_2DFWM_Detuning} }
\end{figure}

\section{Supplementary FWM results}

Here we show supporting experimental and theoretical results concerning the four-wave mixing.
Adding to the data shown in Figs.\,3,4 of the main manuscript, we show here the calculated FWM
dynamics for various detunings in Fig.\,\ref{fig:SM_2DFWM_Detuning}, both the real-time resolved
$|P(t,\tau)|^2$, the frequency resolved $|\tilde{P}(\omega,\tau)|^2$, and the 2D FWM
$|\bar{P}(\omega,\omega_\tau)|^2$.

The phase correction applied to the experimental FWM polarization $\tilde{P}(\omega, {\tau})$
setting the phase evolution of the FWM versus $\tau$ to its expected evolution for an uncoupled
system, is given by

\be \tilde{P}_{\rm cor}(\omega, \tau)= \tilde{P}(\omega, \tau) \exp\left(i\left(\omega_{\rm
cor}\tau-\arg{\tilde{P}(\omega_{\rm cor}, {\tau})}\right)\right). \ee

%2D FWM for T=13.5K:
\begin{figure}
\begin{center}
\iffigs\ifpdf\includegraphics[width=0.6\columnwidth]{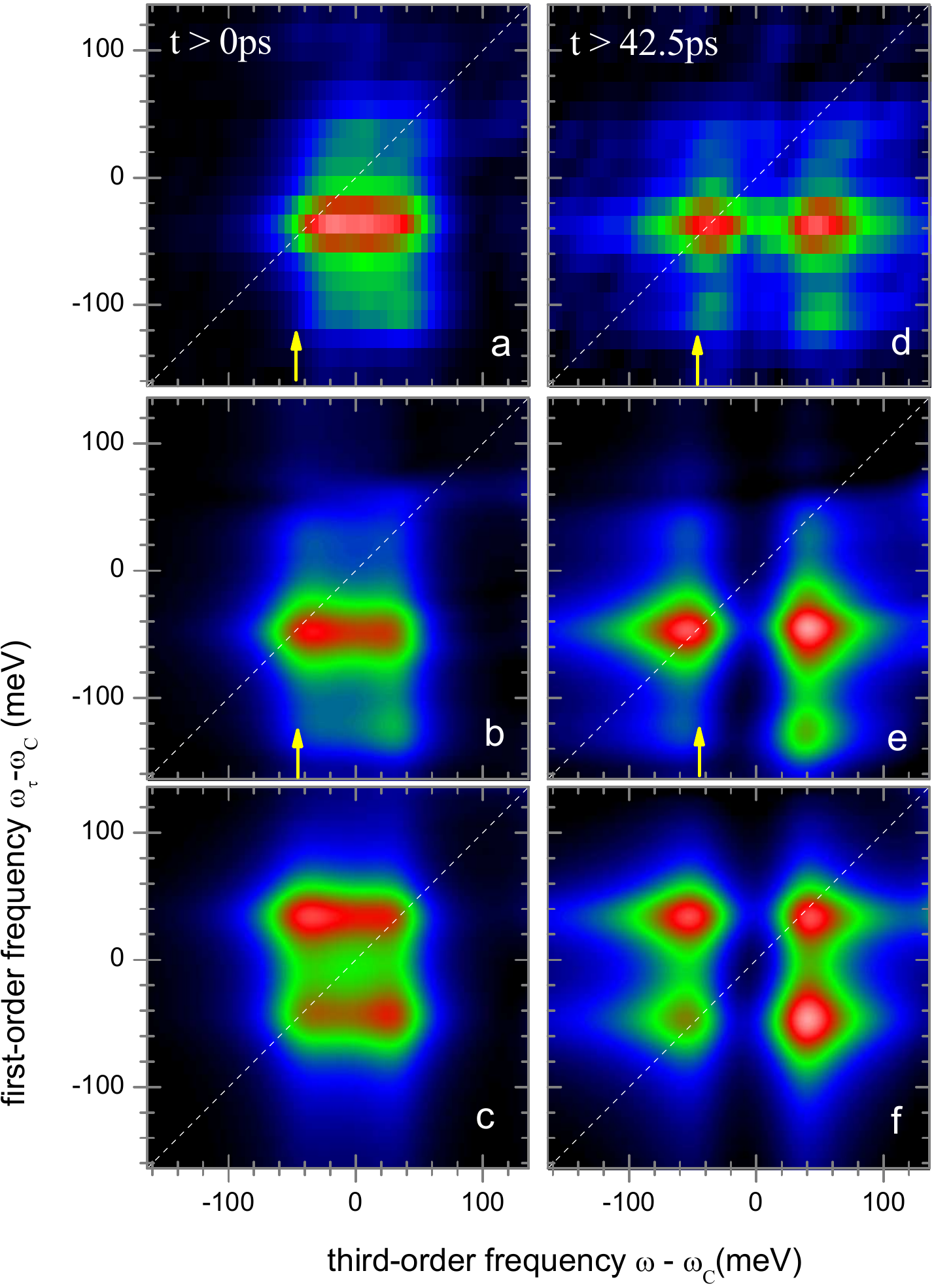}\fi\fi
\end{center}
\caption{ \label{fig:SM_SM_2DFWM_13K} 2D FWM at $T=13.5$\,K ($\delta=-133\,\mu$eV). (a)
$|\bar{P}(\omega, \omega_{\tau})|^2$, phase corrected at $\omega_{\rm cor}=1334.52\,$meV, marked by
the arrow. The diagonal $\omega=\omega_{\tau}$ is shown as dashed line. Predicted data
corresponding to (a) are shown with (b) and without (c) phase correction. In (a-c) a linear colour
scale over 2 orders of magnitude is used. (d) post-selected $|\bar{P}_{\rm
s}(\omega,\omega_\tau,42.5\,{\rm ps})|^2$, and its prediction with (e) and without (f) phase
correction.}
\end{figure}

Coherent coupling analysis for $\delta=-133\,\mu$eV ($T=13.5$\,K) is shown in
Fig.\,\ref{fig:SM_SM_2DFWM_13K}. (a) 2D FWM diagram $|P_{\rm cor}(\omega, \omega_{\tau})|^2$
obtained by phase correcting $P(\omega,\tau)$ at $\omega_{\rm cor}=1334.52\,$meV (as marked by the
arrow) and Fourier-transforming the corrected data along $\tau$. Dash-dotted line represents the
diagonal $\omega=\omega_{\tau}$. The 2D FWM response contains both first- and second-rung signals
masking the coherent coupling features due to the broad spectral width of second- to first-rung
transitions. Theoretical spectra corresponding to (a) are calculated with (b) and without (c) phase
correction. In (a-c) linear colour scale over 2 orders of magnitude is used. (d) 2D FWM retrieved
by using the signal emitted after a time-lag of 42.5\,ps, and the phase correction routine is
applied as in (a). In this case the 28 second- to first-rung transitions are virtually suppressed,
so that the FWM signal is dominated by the 4 first-rung to ground state transitions. An
off-diagonal resonance caused by the cavity mediated coherent coupling of the QDs is detected and
well reproduced by the corresponding theoretical simulation, with (e) and without (f) phase
correction.

\begin{figure}
\begin{center}
\iffigs\ifpdf\includegraphics*[width=0.6\columnwidth]{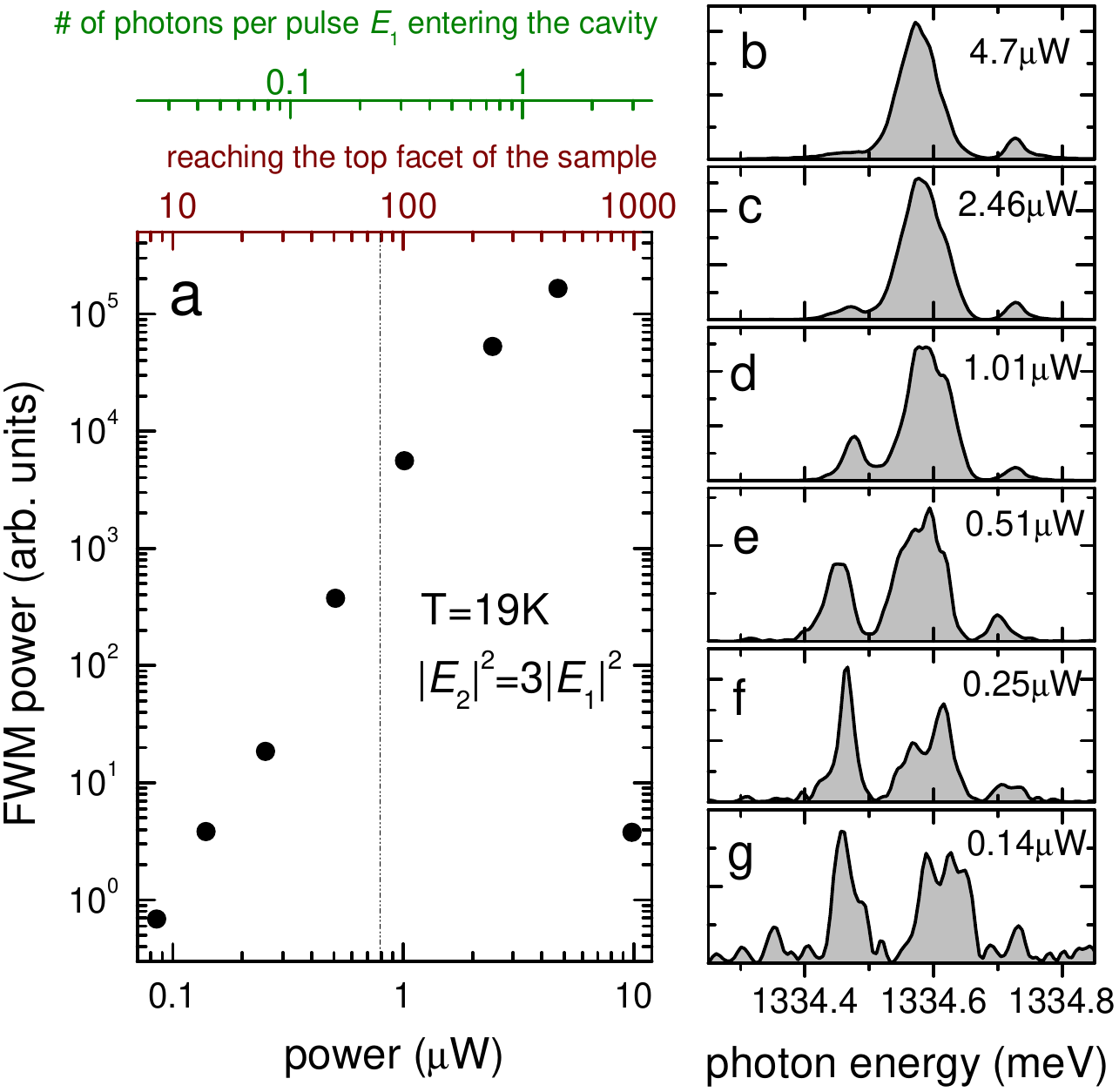}\fi\fi
\end{center}
\caption{ Intensity dependence of the FWM power. (a) Spectrally integrated FWM power as a function
of the combined $E_1$ and $E_2$ driving power, at $\tau=0$ and $\delta=-29\mu$eV (T=19\,K). (b-g)
Spectrally resolved FWM power for different excitation power as labeled.}\label{fig:SM_IntDep}
\end{figure}

To investigate the nonlinear regime of our measurements, we have performed FWM measurements as
function of the excitation power in the pulses $E_1$ and $E_2$ at zero delay time $\tau$. The
resulting FWM spectra and integrated power are shown in Fig.\,\ref{fig:SM_IntDep}.  A third order
scaling is observed, as expected for the lowest order FWM signal, which is proportional to
$E_1^\ast E_2^2$. For higher excitation, the third order scaling saturates, followed by a strong
reduction. The total driving power used in the experiment (marked by the vertical line) is within
the third-order regime. This implies that polariton levels involved in the FWM experiment are
limited to only the first and the second rungs of the TC ladder. We estimate that in the conditions
of our experiment each $E_1$ ($E_2$) pulse delivers 80 (240) photons onto the top facet of the
micropillar. This corresponds, in average, to 0.25 (0.75) photons per pulse that are injected into
the cavity. With increasing power, a transition to a response dominated by the cavity resonance is
observed, resembling a Mollow triplet. We observed an evolution towards a Mollow triplet in a
single X-C system \cite{KasprzakNatureMat10}. Modeling of the high-excitation response will be
reported in a future work.

%\bibliography{paper,langsrv}
\bibliography{MQD_SM}